\documentclass{article}

\date{}
\usepackage{geometry}
\usepackage{multicol,graphicx,amssymb,mathrsfs,amsmath,pifont,amscd,latexsym,amsthm,color,fancyhdr,CJK,url,hyperref,longtable, pifont, subfigure,epstopdf,setspace,multirow,colortbl,tabularx,threeparttable, booktabs, verbatim, bbm,listings, balance}
\usepackage[linesnumbered,boxed,algosection]{algorithm2e}

\begin{document}

\title{Mitigating Redundant Data Transfers for Mobile Web Applications via App-Specific Cache Space}
\author{Yun Ma, Xuanzhe Liu, ShuailiangDong, Yunxin Liu, Tao Xie}

\maketitle

\begin{abstract}
Redundant transfer of resources is a critical issue for compromising the performance of mobile Web applications (a.k.a., apps) in terms of data traffic, load time, and even energy consumption. Evidence shows that the current cache mechanisms are far from satisfactory. With lessons learned from how native apps manage their resources, in this paper, we propose the ReWAP approach to fundamentally reducing redundant transfers by restructuring the resource loading of mobile Web apps. ReWAP is based on an efficient mechanism of resource packaging where stable resources are encapsulated and maintained into a package, and such a package shall be loaded always from the local storage and updated by explicitly refreshing. By retrieving and analyzing the update of resources, ReWAP maintains resource packages that can accurately identify which resources can be loaded from the local storage for a considerably long period. ReWAP also provides a wrapper for mobile Web apps to enable loading and updating resource packages in the local storage as well as loading resources from resource packages. ReWAP can be easily and seamlessly deployed into existing mobile Web architectures with minimal modifications, and is transparent to end-users. We evaluate ReWAP based on continuous 15-day access traces of 50 mobile Web apps that suffer heavily from the problem of redundant transfers. Compared to the original mobile Web apps with cache enabled, ReWAP can significantly reduce the data traffic, with the median saving up to 51\%. In addition, ReWAP can incur only very minor runtime overhead of the client-side browsers.
\end{abstract}

\section{Introduction}
\label{sec:introduction}
Redundant transfer of resources\footnote{In this paper, resources refer to resource objects constituting an app (such as HTML, JavaScript, CSS, and images of a Web app; native code, media files, and layout files of a native app)} refers to the case where a previously fetched resource is downloaded again from the network before the resource is actually updated. For mobile Web applications (a.k.a. apps)~\cite{serrano:2013}, redundant transfers remain as a critical performance issue leading to duplicated data transmission, long page load time, and high energy drain~\cite{Qian:MobiSys2012}\cite{Wang:NSDI2013}.

Redundant transfers originate from apps' resource-management mechanism that is to determine whether a resource should be loaded locally or remotely. Web cache is a conventional resource-management mechanism adopted by Web apps. Web developers can configure cache policies, such as expiration time and validation flag, on those resources that are likely to be loaded from the local storage. The browser maintains a cache space and deals with the cache logic for all the Web apps running in it. However, our previous work~\cite{Ma:WWW2015} found that there is a big gap between the ideal and actual cache performance of mobile Web apps. For example, for the mobile versions of top-100 websites of Alexa, although more than 70\% of resources can be loaded from the cache when these websites are revisited after one day, less than 50\% of these cacheable resources are actually loaded from the cache. Surprisingly, all resource transfers are redundant for some well-known websites when they are revisited after one day. We also revealed two major causes for redundant transfers: (1) the imperfect cache configuration, such as heuristic expiration and conservative expiration time; and (2) the undesirable Web development practice, such as using random strings to name resources for enforcing their refresh.

Due to the dynamics of mobile Web apps, it is difficult for Web developers to properly configure the apps' cache policies. Short expiration time may lead to redundant transfers, while long expiration time may result in the usage of stale resources. As a result, using cache policies is not a desirable mechanism to accurately determine whether resources should be loaded locally or remotely~\cite{Liu:TMC2016}.

To fundamentally reduce redundant transfers for mobile Web apps, the resource management in native apps can provide some useful inspirations. Resources of native apps are managed directly and explicitly by app-specific logics to control where to load resources and when to update the local resources. Intuitively, native apps explicitly distinguish their static resources from dynamic ones, and encapsulate the static resources into a package that is installed into a dedicated space allocated by the underlying operating system. When a native app is running, the app logic controls that only its dynamic resources are downloaded on demand to provide the ¡°fresh¡± data to users, while the static resources in the installed package are always fetched locally. When the static resources have to be updated, a new resource package is downloaded and installed to refresh all the static resources.

However, there are two main challenges for Web developers to adopt such a package-based resource management specific to a Web app. First, it is hard to maintain the resource package. Resources of Web apps are loosely coupled, and usually updated independently and casually without influencing each other. As a result, it is tedious and error-prone to decide which resources should be put into the package and when to update the package. Second, it is hard to enable mobile Web apps to use the package. Modern Web apps are complex and there are many mature Web development frameworks. As a result, a lot of manual efforts have to be needed to realize or refactor mobile Web apps to benefit from the package-based resource management.

To address these challenges, in this paper, we propose the \emph{ReWAP} approach to restructuring mobile Web apps to be equipped with package-based resource management while requiring minimal developer efforts. The key rationale of ReWAP is to provide more efficient and app-specific control of resource management rather than relying on only the current mechanisms such as Web cache, to avoid the caused unnecessary redundant resource transfers. By retrieving the update of resources of mobile Web apps, ReWAP automatically maintains resource packages by accurately identifying which resources should be loaded from the local storage for a considerably long period. Based on the package information, ReWAP automatically checks the update of resource packages, refreshes the resources in the package when the resource package is updated, and loads resources from resource package for mobile Web apps. To integrate ReWAP with existing mobile Web apps, Web developers need only minor modifications to their existing apps. In summary, \emph{ReWAP shares the same spirit of resource management mechanisms as those in the installation package of native apps but in the way of Web.}

To the best of our knowledge, our work is the first to facilitate Web developers to effectively reduce redundant transfers of mobile Web apps by conducting resource management in a similar way as native apps. More specifically, this paper makes the following main contributions:

\begin{itemize}
    \item{We design ReWAP, a packaging approach for a mobile Web app to accurately identify resources that can be loaded from the local storage for a considerably long time. The maintained package can maximize the benefit of data-traffic saving by considering all the users of the mobile Web app.}
    \item{We implement ReWAP to minimize developer efforts of restructuring existing mobile Web apps. Web developers can easily integrate ReWAP in their current implementation of a mobile Web app, and the end-users are completely unaware of the existence of ReWAP when they access the Web apps.}
    \item{We conduct experiments based on 15-day access logs of 50 mobile Web apps that suffer heavily from redundant transfers to evaluate the effectiveness of ReWAP. Compared to the original mobile Web apps with cache enabled, ReWAP can significantly save the data traffic with the median value up to 51\% and the maximum of almost 100\%. In addition, ReWAP incurs only quite small runtime overhead of the client-side browsers.}
\end{itemize}

The remainder of this paper is organized as follows. Section~\ref{sec:example} illustrates the problem of redundant transfers with an example and compares the resource-management mechanisms of Web apps and native apps. Section~\ref{sec:overview} presents the overview of the ReWAP approach. Sections~\ref{sec:package} and~\ref{sec:wrapper} show the details of ReWAP's key components, i.e., Package Engine and Wrapper, respectively. Section~\ref{sec:implementation} describes the implementation of ReWAP and demonstrates its easy deployment. Section~\ref{sec:evaluation} evaluates ReWAP based on top Web apps of Alexa. Section~\ref{sec:related} discusses the related work and Section~\ref{sec:conclusion} concludes the paper.

\section{Background and Motivation}
\label{sec:example}
In this section, we present the background and motivation for leveraging the resource-management mechanisms of native apps to improve Web apps. We first describe a motivating example to illustrate the problem of redundant transfers. Then we compare the resource-management mechanisms of Web and native apps.

\subsection{Redundant Transfer in Mobile Web Apps}
\begin{figure*}
\centering
\begin{center}
\includegraphics[width=1\textwidth]{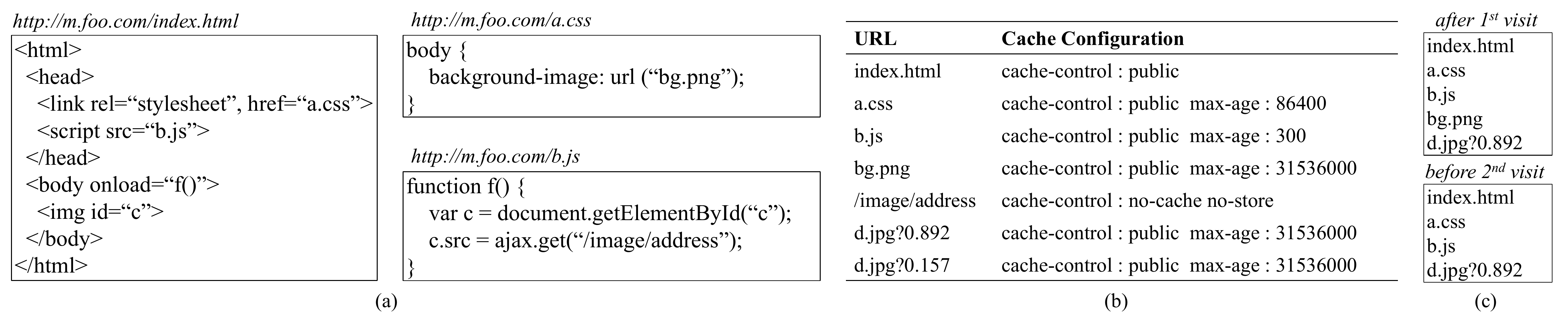}
\caption[7.5pt]{Motivating example. (a) Resource excerpts of a Web app; (b) Cache configuration; (c) Cache entries between two visits.}\label{fig:example}
\end{center}
\end{figure*}

Although the resource management of Web apps is flexible enough to achieve easy-to-access and always-updated features, it could lead to redundant transfers of resources. We illustrate redundant transfers via an example shown in Figure~\ref{fig:example}.

Figure~\ref{fig:example}(a) are resource excerpts of a mobile Web app ``http://m.foo.com/''. The HTML resource indicates that the app includes a layout resource ``a.css'' and a JavaScript resource ``b.js''. When ``a.css'' is being evaluated, a background image ``bg.png'' is identified. After parsing the HTML resource is finished and the \texttt{onload} event is triggered, JavaScript function \texttt{f} is executed to get an address of an image by requesting a service ``/image/address''. Suppose that the returned address is ``d.jpg?0.892'', and then the image is retrieved. Therefore, when the app is visited at the first time, 6 resources are actually retrieved. Figure~\ref{fig:example}(b) shows the cache configuration of these resources. The HTML is not configured with an explicit expiration time so the browser assigns a random time that is usually not very long, e.g., 30 minutes. The expiration time of CSS, JavaScript, and images are configured as 1 day, 5 minutes, and 1 year, respectively. The service ``/image/address'' is configured as no-cache and no-store to ensure obtaining the latest address at every visit. The top table in Figure~\ref{fig:example}(c) shows the resources in the browser cache after the first visit.

Assume that the app is revisited after one hour and all the related resources have not been updated except the ``/image/address''. The bottom table in Figure~\ref{fig:example}(c) shows the cached resources before the second visit. It can be seen that the background image ``bg.png'' has been removed out of the cache due to the limited size of cache on mobile devices because all the Web apps accessed by a browser share a fixed size of cache space.

Given the current status of cache, at the revisit of this app, several resources that could have been loaded from the cache are actually re-downloaded from the network, leading to redundant transfers of resources falling in the following main categories~\cite{Qian:MobiSys2012,Ma:WWW2015}.

\noindent \textbf{RT1: Resources that are moved out of the cache.}

Due to the imperfect implementation of cache on mobile browsers such as limited size and non-persistent storage, resources in the cache may be removed out of the cache after some time. In the preceding example, the background image ``bg.png'' is removed out of the cache and has to be re-downloaded when the Web app is revisited.

\noindent \textbf{RT2: Resources that are wrongly judged as expired.}

Each resource has to be configured by developers with a cache policy. Due to the imperfect cache configuration of resources whose expiration time is either configured to be too short or not configured but assigned heuristically by browsers, many resources are falsely judged by browsers as expired ones, and have to be validated or re-downloaded. In the example, the HTML resource has not been assigned an explicit expiration time, and the expiration time of the JavaScript resource is configured to be too short. As a result, these two resources cannot be loaded from the local environment when the Web app is revisited.

\noindent \textbf{RT3: Resources that are requested by new URLs but have the same content with cached ones.}

Resource Loader of browsers uses URLs to uniquely distinguish among resources. Resources with different URLs are regarded as totally different ones. In the example, the same image ``d.jpg'' has different URLs at two visits, resulting in being fetched twice. URL changing is usually adopted to realize backend load balance according to URL routing. Although such a practice can improve the performance of backend servers, it actually harms the loading process of mobile Web apps.

\subsection{Resource Management of Web Apps and Native Apps}
\begin{figure}
\centering
\begin{center}
\includegraphics[width=0.7\textwidth]{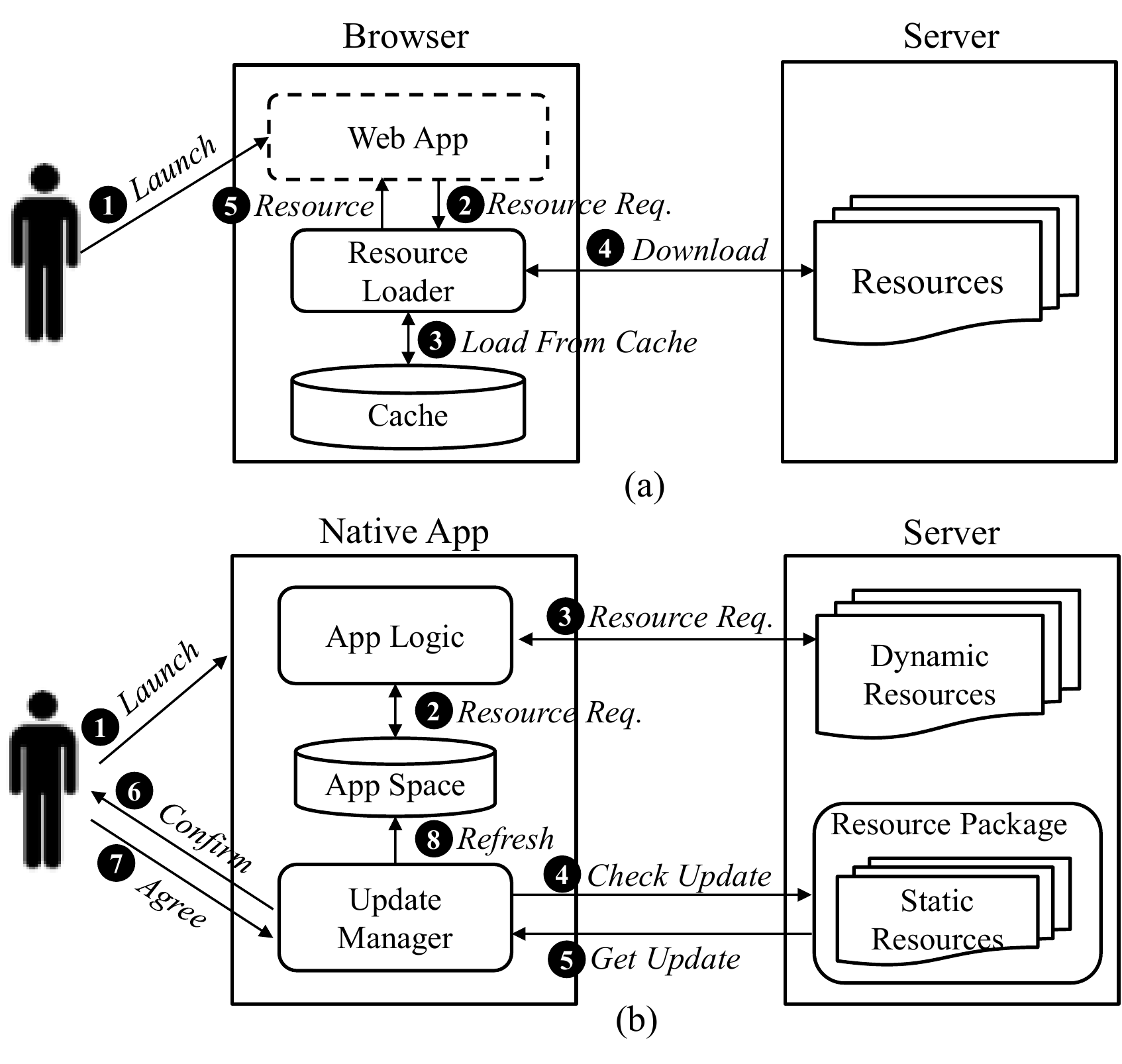}
\caption[7.5pt]{Resource management of (a) Web apps and (b) native apps.}\label{fig:apparchi}
\end{center}
\end{figure}
A key reason for the preceding redundant transfers is the inefficient resource management of Web apps. Figure~\ref{fig:apparchi}(a) illustrates the resource-management mechanism of Web apps. Web apps rely on the underlying browsers to manage their resources. All Web apps in a browser share a common cache space whose size is usually small on mobile devices. When a user launches a Web app in the browser (\textcircled{\small{1}}), resources for rendering the Web app are dynamically identified and all the resource requests are handled by the Resource Loader component in the browser (\textcircled{\small{2}}). Based on the Web cache mechanism~\cite{RFC2616}, the Resource Loader determines whether to load the resource from the cache (\textcircled{\small{3}}) or downloads it from the server (\textcircled{\small{4}}). After retrieving the resource, the Resource Loader returns it to the Web app (\textcircled{\small{5}}). In summary, Web apps rely on the app-independent browser logics to manage resources. Such a mechanism makes Web apps flexible for resource management so that Web apps can be always up-to-date. However, Web apps cannot have the full control of resources to be loaded from the local storage and when to update the local resources. As a result, redundant transfers arise when the cache policies are not configured properly or the browser removes cached resources.

In contrast, the resource management of native apps works in a different fashion and can be more efficient. Figure~\ref{fig:apparchi}(b) illustrates the resource-management mechanism of native apps. Native apps separate resources into two sets, i.e., the dynamic resource set and the static resource set. Static resources are encapsulated into a resource package. Before using the native app, the resource package has to be installed on the device. When a user launches the app (\textcircled{\small{1}}), the App Logic controls to load static resources from the App Space (\textcircled{\small{2}}) and dynamic resources from the server (\textcircled{\small{3}}). Usually, there is a built-in Update Manager for updating the static resources. The Update Manager checks update with the server (\textcircled{\small{4}}) in some situations (e.g., every time when the app is launched) to find whether the resource package has been updated (\textcircled{\small{5}}). If a new package is retrieved, the Update Manager confirms with the users whether to update the app (\textcircled{\small{6}}). If agreed (\textcircled{\small{7}}), then the Update Manager refreshes the static resources with the new resource package (\textcircled{\small{8}}). In summary, native apps have app-specific logics to control the resources loaded from the local environment and the update of local resources.

Comparing the two resource-management mechanisms reveals that native apps can manage their resources based on app-specific logic with resource packages while Web apps cannot precisely manage their resources. The insight underlying our new approach is that redundant transfers originate from the principle of the resource-management mechanism adopted by Web apps.

\section{Approach Overview}
\label{sec:overview}
To fundamentally reduce the redundant transfers, we propose our new solution with the key rationale of lessons learned from the resource-management mechanism used by native apps, while keeping the advantages of the mechanism used by Web apps. More specifically, mobile Web apps can encapsulate stable resources into a package and make the resources in the package always loaded locally rather than being fetched from the servers, while other resources are regularly loaded by the browser's default mechanism. All the resources in the package are refreshed together also by the default mechanism only when the resource package get updated. The update of package should also follow the way of the Web without the intervention of end users.

To this end, we propose the ReWAP approach to restructuring mobile Web apps to be equipped with package-based resource management. ReWAP can accurately identify the resources that should be loaded from the local storage for a considerably long time and that can be refreshed together with minimal cost when the package is updated. Other than the native apps, such a packaging mechanism follows the conventional way of the Web, i.e., the updating and refreshing of packaged resources still use the browser's default cache mechanism. The Web developers can simply integrate ReWAP into their existing mobile Web apps with only minor modifications. As we will show later, they need only to redirect the entrance of the app to a Wrapper that delegates the resource loading. Meanwhile, the client-side browser performs regularly without additional modifications.

\begin{figure}
\centering
\begin{center}
\includegraphics[width=0.7\textwidth]{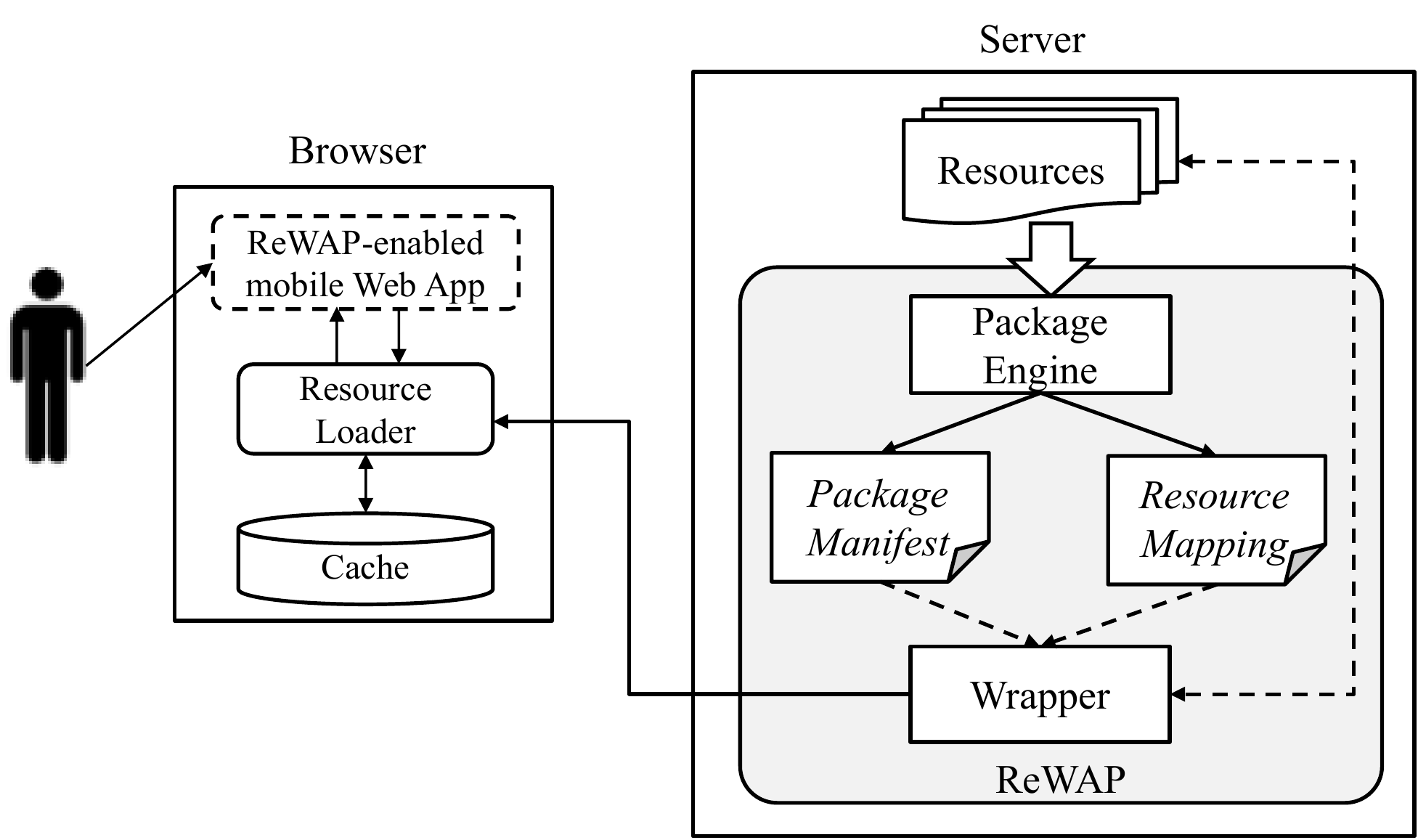}
\caption[7.5pt]{The ReWAP approach.}\label{fig:overview}
\end{center}
\end{figure}

Figure~\ref{fig:overview} shows the overview of the ReWAP approach. In general, ReWAP can be deployed at the server of Web apps. It consists of a Package Engine to automatically generate and maintain resource packages, and a Wrapper for mobile Web apps to use and update resource packages in the local storage. Each Web app has a unique resource package for all their users, just like the installation package of a native app. Note that a mobile Web app can consist of many Web pages. Our approach currently treats each Web page independently. As different Web pages of a Web app may share similar resources~\cite{Wang:WWW2012}, it is possible to further improve the performance by packaging all the Web pages of a Web app together. Such optimization is to be figured out as future work. In the rest of this paper, the term ``Web app'' denotes a single Web page referenced by an HTML document.

By retrieving the update of resources constituting a mobile Web app, the Package Engine generates and maintains a resource package with two configuration files: Package Manifest and Resource Mapping. The Package Manifest specifies which resources are in the resource package. The update of Package Manifest indicates the update of the corresponding resource package. The Resource Mapping keeps the relationship between URLs and unique resource entities. Resources that have the same content but are identified by different URLs are mapped into one resource entity according to Resource Mapping. Therefore, the generated package is highly accurate to cover more resources.

The Wrapper is essentially a separate HTML page where Web developers can easily enable their mobile Web apps with the package-based resource management. When a ReWAP-enabled mobile Web app is launched, the Wrapper is first fetched from the server. Then the Wrapper controls the loading process on the browser (we use dotted lines to represent the flow taking place on the client side). The Wrapper checks whether the resource package has been updated according to the Package Manifest. If updated, all resources in the package are refreshed and stored into an App-Specific Space according to Resource Mapping. After the refreshing, the Wrapper loads the app. The Wrapper intercepts all the resource requests to determine whether to load a resource from the App-Specific Space or as usual based on the Package Manifest.

ReWAP is deployed as a service on the same server with the target mobile Web app. For example, to integrate ReWAP with the motivating mobile Web app in Section~\ref{sec:example}, the developer can specify the app's URL http://m.foo.com/index.html and then launches ReWAP service on the server m.foo.com. The Package Engine is then automatically started as a background process at the server side, while the Wrapper is also generated on the server with a URL, e.g., http://m.foo.com/index/wrapper.html. At last, the developer configures the server, making the requests to index.html redirected to the URL of the Wrapper. When a user visits the ReWAP-enabled app, the Wrapper is loaded first to the browser, dealing with resource packages. Then the Wrapper loads the index.html by an AJAX call and intercepts all the resource requests. On the whole, deploying ReWAP requires only minimal modifications to existing mobile Web apps.

In the next two sections, we present the technical details on how the Package Engine maintains the resource package and how the Wrapper supports resource package at the runtime of Web apps.

\section{The Package Engine}
\label{sec:package}
\begin{figure}
\centering
\begin{center}
\includegraphics[width=0.6\textwidth]{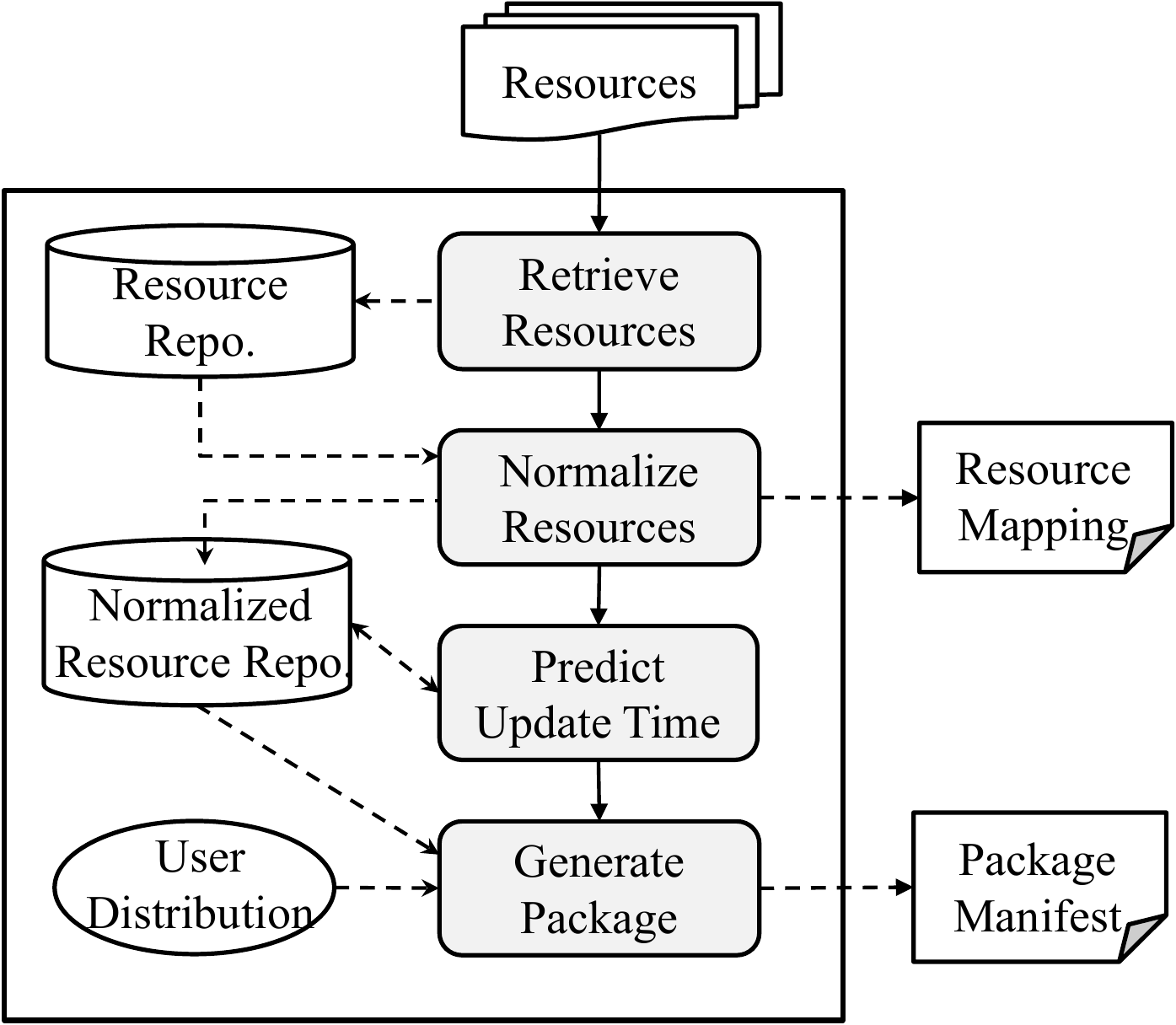}
\caption[7.5pt]{The workflow of the Package Engine.}\label{fig:package}
\end{center}
\end{figure}
Figure~\ref{fig:package} shows the four phases to maintain the resource package. In the first phase \emph{Retrieve Resources}, the Package Engine repeatedly visits the Web app via a real browser runtime (e.g., WebKit) that is deployed at server-side, retrieves all the needed resources to render the app, and stores them into the Resource Repository. The visit is performed with a given time interval or every time when informed by the server. In the second phase \emph{Normalize Resources}, based on the resources in the Resource Repository, resources that have the same content but identified with different URLs are normalized as a unique resource that is not required to be re-downloaded. The normalized resource is identified based on a regular expression. Resources whose URLs can match the regular expression are treated as the same resource. A Resource Mapping file is generated to maintain the relationship between normalized and original resources. In the third phase \emph{Predict Update Time}, according to the history of resource updates kept in the Normalized Resource Repository, the Package Engine predicts the update time of each normalized resource. In the last phase \emph{Generate Package}, a subset of normalized resources is selected to generate the Package Manifest based on the benefit brought to the mobile Web app. Since all the users of an app share the same package, we use the averagely saved data traffic as the main metric to quantify the benefit. The generation takes place in two conditions. If the current resource package is invalid, i.e., any resource in the package has been changed, then a passive selection is performed to select the resources with the highest benefit to generate the new resource package. Otherwise, if the current package is still valid, an active selection can generate a new package only when the benefit of the new package exceeds the old one over a given threshold. We next present the details of each phase, respectively.

\subsection{Retrieving Resources}
In the first phase, by periodically revisiting the target Web app, we can retrieve the update of the Web app's resources. We define a ``\emph{concrete resource}'' structure to represent each retrieved resource. A ``concrete resource'' has the following fields: (1) \emph{URL}, which is the identifier of the resource; (2) \emph{MD5}, which is the checksum of the resource content; (3) \emph{size}, which is the length of the resource content; (4) \emph{cache duration}, which is the configured expiration time. All these fields can be obtained from the HTTP response message. We define $R_t$ as the set of concrete resources retrieved at time $t$.

In order to ensure the completeness of the retrieved resources, we employ a real browser facility to actually launch and render the Web app. Otherwise, a lot of resources may be missed by only parsing the HTML document. We record all the HTTP traffic occurred by the browser facility in the progress of visiting a Web app.

Here, an important issue is the timing to retrieve resources because we have to ensure that the resource status is always up-to-date to avoid inconsistencies. One possible solution is to make the server notify the Package Engine at the time when the resources are updated so that resource updates can be exactly captured. Another solution is to trigger by a retrieving frequency. The retrieving frequency can be self-adjusted according to the features of the target Web apps. For example, apps that change more often can be revisited more frequently. In our current design, we use a fixed retrieving frequency to trigger resource retrieving. We leave the server-informed mechanism and self-adjusted frequency to the future work.

\subsection{Normalizing Resources}
In the second phase, the Package Engine identifies the resources that have different URLs but the same content at different visits to the Web app. We denote this kind of resources as ``\emph{cross-dress resources}''. We normalize cross-dress resources into one normalized resource. We keep the relationship of the normalized resource and one concrete resource in the Resource Mapping. The final Package Manifest consists of normalized resources so that cross-dress resources do not have to be re-downloaded for multiple times.

We observe that there are two frequent URL patterns of the cross-dress resources. One is the query strings generated by JavaScript, e.g. \emph{Math.random()}, or by server scripts. In the motivating example, the URL of the image ``d.jpg'' has two different query strings ``?892'' and ``?157'' at two visits but the image does not change. In such a case, the URLs vary only in the query, i.e., the random value. The other pattern is the \emph{CDN prefixes}. At different visits, resources could be retargeted at different CDN servers. In such a case, the paths of the URL are the same but the domain part of the URL could be changed according to the target CDN servers. Based on the two patterns, we assume that the URLs of cross-dress resources can be different in a certain part of the URL. Therefore, we can apply the \emph{Longest-Substring} algorithm to find the base string of the URLs and use a regular expression to represent the changing part. For example, the image   ``d.jpg'' in the motivating Web app has two different query strings ``?892'' and ``?157'' at two visits. So we can use the regular expression ``$d.jpg\backslash?*$'' to represent the normalized image.

We define a ``\emph{normalized resource}'' structure to represent a unique normalized resource. A normalized resource is generated by aggregating cross-dress resources. It has all the fields of the ``\emph{concrete resource}'' structure. The additional fields of ``\emph{normalized resource}'' include: (1) \emph{expression}, which describes the URL pattern of cross-dress resources; (2) \emph{predicted time}, which describes the estimated duration time that the resource remains unchanged; and (3) \emph{status}, which records all the historic statuses of the resource. We use $status_t$ to denote the status at time $t$. Each statust can be ``\emph{inexistent}'', ``\emph{changed}'', or ``\emph{unchanged}''. Such historic status information is used to predict the update time. We define $H_t$ as the set of normalized resources at time $t$. $H_t$ is updated every time when concrete resources are retrieved.

We should enforce a one-to-one mapping between the normalized resources and the concrete resources of each visit in order to prevent resources with different contents from being matched to one normalized resource. To this end, we compare the MD5 of the resource content to determine whether two resources are the same.

\begin{algorithm}
    \caption{Normalize resources.}
    \label{algo:normalize}
    \SetAlgoLined
    \KwIn{Last set of normalized resources $H_{t-1}$, current set of concrete resources $R_t$}
    \KwOut{Updated set of normalized resources $H_t$}
        INITIAL $H_t\leftarrow H_{t-1}$;\\
        \ForEach{$h\in H_t$} {
            INITIAL $h.status_t\leftarrow ``inexistent"$;
        }
        \ForEach{$r\in R_t$} {
            $P\leftarrow FindSameURL(H_t, r)$;\\
            $q\leftarrow FindSameMD5(H_t, r)$;\\
            \If{$q\neq null$} {
                $q.expression\leftarrow CalRegExpr(q.expression, r.URL)$;\\
                $q.status_t\leftarrow ``unchangecd"$;
            }
            \ElseIf{$P.size = 1$} {
                $P.status_t\leftarrow ``changed"$;\\
                $UpdateResource(P)$;
            }
            \Else{
                $RemoveResource(P)$;\\
                $AddResource(r)$;
            }
        }
        $CheckMapping(R_t, H_t)$;\\
        return $H_t$.
\end{algorithm}

Algorithm~\ref{algo:normalize} describes the process of managing the set of normalized resources. Given the last set of normalized resources $H_{t-1}$, and the current set of concrete resources $R_t$, the algorithm returns the updated set of normalized resources $H_t$. We assign $H_{t-1}$ to $H_t$ at first and initialize all resources' status of time $t$ as ``\emph{inexistent}'' (Lines 1-4). Then we add, update, and remove resources in different cases (Lines 5-20). Additionally, we need to handle conflicts in the new set of normalized resources to ensure the one-to-one mapping (Line 21).

\subsection{Predicting Update Time of Resources}
In the third phase, we infer whether a resource is sufficiently stable by predicting the update time of the resource. We assume that the evolution history of a resource can reflect the trend of resource updates. For example, if a resource is updated every day in the history, it is likely to be updated in the next day. Therefore, we design an algorithm to predict the update time of resources based on their evolution histories.

Algorithm~\ref{algo:predict} shows the details of how our prediction works. By examining the evolution histories of some resources, we find that after a resource disappears at one visit, the possibility of its reappearance is rather small. So every time when an ``\emph{inexistent}'' status is captured for a resource, we immediately set its predicted time to 0 (Lines 1-3). For other resources, we mainly capture the total times of ``\emph{changed}'' status and predict the next time when the resource is likely to change. In some cases, the resource can update in an unusual fashion, so we should not aggressively change the predicted time. For example, if a resource is updated once every day in the history and at one time it is updated one hour after the last update, we should use a modest way to reduce the predict time. Here, we use the gradient descent algorithm~\cite{Zhang:ICML2004} to predict the resource update time (Line 5). Furthermore, if no ``changed'' status is captured, we set the predicted time according to the numbers of ``unchanged'' other than infinite (Lines 6-8). Finally, we remove all the resources whose predicted time is 0 in order to limit the numbers of historic resources (Lines 10-12).
\begin{algorithm}
    \caption{Predict update time of normalized resources.}
    \label{algo:predict}
    \SetAlgoLined
    \KwIn{Historic status $status_0,\dots,status_t$ of a normalized resource $h\in H_t$, visiting interval $vi$}
    \KwOut{Predicted update time of $h$}
        \If{$h.status_t = ``inexistence"$} {
            $h.predictedtime\leftarrow 0$;
        }
        \Else{
            $h.predictedtime\leftarrow GDM(status_0,\dots, status_t)$;\\
            \If{$h.predictedtime = inf$} {
                $h.predictedtime\leftarrow |status.unchanged| * vi$;
            }
        }
        \If{$h.predictedtime = 0$} {
            $RemoveResource(h)$;
        }
\end{algorithm}

\subsection{Generating Package}
After estimating the update time of each resource, we can select the stable resources that can be packaged. We regard that resources in the package can bring some benefit to the Web app. We assume that all the Web app's users share a common resource package. Given that different users may revisit the Web app at different times and with different frequencies, the benefit varies among different users. To measure the overall benefit that can be gained by the packaged resource, we assume a user distribution function $\sigma$ to represent the percentage of users at different revisiting intervals. We define a metric, the average saved data traffic, to quantify how much data traffic all the Web app's users can save on average given the user distribution function.

Let us assume that a subset of resources $M\subset H_t$ is selected. Suppose that $T$ is the minimum predicted time in $M$, we then have an expectation that such a resource package $M$ will be updated at time $T$. For each normalized resource in $M$, the browser does not need to request this resource before $T$. On the contrary, without the resource package, each resource should be requested from the server after its cache duration. Thus, for each resource, if $T$ exceeds the cache duration, the traffic saving comes from the difference between the configured cache duration and our predicted time. If the predicted time is less than the cache duration, the resource can still be loaded from the cache according to our package mechanism, incurring no extra traffic.

Algorithm~\ref{algo:select} shows how to select the best resource package. We first sort the normalized resources according to the predicted time in ascending order (Line 1). Among all subsets of resources whose minimum predicted time is $T$, the benefit of a smaller set cannot exceed that of a bigger one. Thus we do not need to enumerate all potential packages. We enumerate the potential $T$ (Lines 2-10), and calculate the benefit of the largest set whose minimum predicted time is $T$ (Lines 5-10). Finally, we choose the package that provides maximum benefit (Lines 11-13).

\begin{algorithm}
    \caption{Select the packaged resources.}
    \label{algo:select}
    \SetAlgoLined
    \KwIn{Current set of normalized resources $H_{t}$, user distribution $\sigma$}
    \KwOut{Resource package $M$}
        Sort $H_t$ based on its predicted time in ascending order;\\
        \For{$i\leftarrow 0~to~|H_t|$} {
            $benefit(i)\leftarrow 0$;\\
            $T\leftarrow H_i.predictedtime$;\\
            \For{$j\leftarrow i~to~|H_t|$} {
                \If{$H_j.cacheduration < T$} {
                    $benefit(i) += \sigma(H_j.cacheduration, T) * H_j.size$;
                }
            }
        }
        Select $i$ where $benefit(i)$ is the largest;\\
        $M\leftarrow H_t(i, i+1, \dots, |H_t|)$;\\
        return $M$.
\end{algorithm}

To make the resource package sufficiently stable, we may not always use the resource package with the largest benefit. We first check whether the latest resource package is still valid where all the resources in the latest package has not been updated. If any resource is changed, we just use the package generated by Algorithm~\ref{algo:select} to replace the invalid package. If the latest resource package is still valid, then we replace the latest package only when the benefit of the new package excesses the current one by a given threshold.

\section{The Wrapper}
\label{sec:wrapper}
\begin{figure}
\centering
\begin{center}
\includegraphics[width=0.6\textwidth]{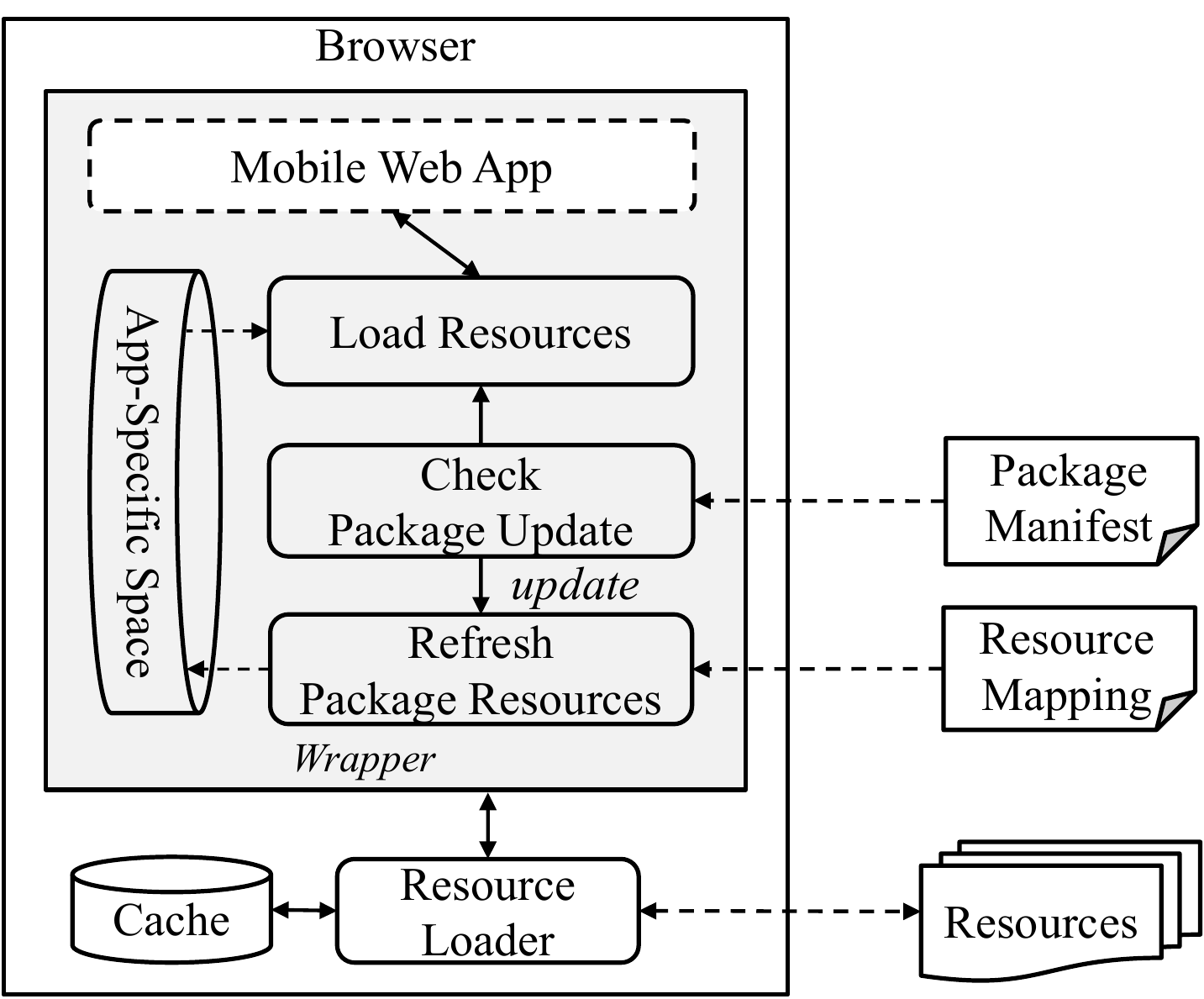}
\caption[7.5pt]{The workflow of the Wrapper.}\label{fig:wrapper}
\end{center}
\end{figure}
The functionality of the Wrapper is to equip mobile Web apps with the ability to use the resource packages atop the resource-management mechanism of Web browsers. Figure~\ref{fig:wrapper} shows the workflow of the Wrapper. The Wrapper has an App-Specific Space, which is a dedicated local storage to store packaged resources for each Web app. Each Web app has its own App-Specific Space that is not shared with other Web apps.

When an end user visits the ReWAP-enabled mobile Web app, the Wrapper is first loaded from the server and runs in the browser. It has three phases to load the target mobile Web app.

\textbf{Checking Package Update.} After the Wrapper is loaded, it checks whether the corresponding resource package has been updated. The Wrapper communicates with the Package Engine to check whether the previously retrieved Package Manifest has been updated. If not updated, then there is no need to refresh the local resources and the Wrapper starts to load the target Web app. Otherwise, if the package has been updated, the Wrapper refreshes the packaged resources stored in the App-Specific Space. Only after the App-Specific Space has finished refreshing, can the target mobile Web app start loading in order to ensure all the resources are up-to-date.

\textbf{Refreshing Package.} When the packaged resources have to be updated, the Wrapper refreshes all the resources together according to the newly retrieved Package Manifest. The update process follows the regular Web resource loading mechanism that each resource request is handled by the Resource Loader of the browsers. For normalized resources, the corresponding concrete resource is requested according to the Resource Mapping.

\textbf{Loading Resources.} When the package updates finishes, the Wrapper loads the target Web app. While loading, the Wrapper intercepts all the resource requests emitted by the Web app. For each resource request, the Wrapper checks the App-Specific Space by matching the URL with regular expressions specified in the Package Manifest. If found, then the resource is directly returned to the app from the App-Specific Space. Otherwise, the request is forwarded to the Resource Loader of browsers to retrieve the corresponding resource either from the cache or from the server.

Note that the update check of resource packages and the refresh of resources are actually handled by the Resource Loader of the client-side browsers to manage the resource package with the regular Web mechanism.

\section{Implementation}
\label{sec:implementation}
We implement ReWAP by leveraging HTML5~\cite{HTML5} Application Cache interface (in short as AppCache). AppCache is designed for offline Web apps, but it actually provides a way for developers to control their resources. Since HTML5 has already been supported by most of the popular commodity mobile browsers such as Chrome, Safari, and Firefox, mobile Web apps with ReWAP can run directly on the latest mobile browsers. Therefore, we can realize the easy and fast deployment without introducing extra cost to end users.

In this section, we first give some background knowledge of HTML5 Application Cache and then describe the details of our implementation and deployment.

\subsection{HTML5 Application Cache}
The Application Cache (AppCache) is a feature of HTML5 that aims to allow Web apps to be reliably accessed when the browser is offline. To enable AppCache, developers provide a manifest file to specify what resources are needed for Web apps to work offline, and configure the manifest file to the``\emph{manifest}'' attribute of an HTML document's \texttt{<html>} tag. The manifest file mainly consists of two sections. Each section has a list of URLs specifying the behavior of the corresponding resources.
\begin{itemize}
    \item{\textbf{CACHE.} Resources listed in this section are explicitly cached after they are downloaded for the first time. Even when the browser is online, these resources are still loaded from the AppCache rather than being downloaded from the network. Note that HTML documents referring to manifest files are set in the CACHE section by default.}
    \item{\textbf{NETWORK.} Resources listed under this header can bypass the AppCache and be requested regularly by the browser. When the browser is offline, these resources cannot be loaded from the AppCache. A wildcard flag ``*'' can be used to make any resource that is not listed in the CACHE section bypass the AppCache mechanism.}
\end{itemize}

\subsection{Implementation of the Package Engine}
Based on the AppCache specification, the concept of Package Manifest in ReWAP can be implemented by the manifest file of AppCache. However, as the AppCache manifest can use only concrete URLs, we put the concrete resources in the manifest and use the generated Resource Mapping file to help check whether a resource is in the package or not by matching with the Resource Mapping. We implement the Package Engine in Java. The Package Engine can be published as a stand-alone component and deployed as a service at the server-side. For the browser facility of the Package Engine, we use the Chromium Embedded Framework~\cite{CEF} to render the mobile Web apps and record the HTTP traffic.

\subsection{Implementation of the Wrapper}
The Wrapper is totally implemented as an HTML page with AppCache enabled together with a JavaScript library. The Wrapper is totally implemented by standard Web technologies so that it can run directly on modern mobile browsers.

The HTML page is configured to use the manifest file generated by the Package Engine. The App-Specific Space and Check Package Update can be provided by the AppCache. To realize loading the target Web app, the page registered a JavaScript callback function on the \texttt{onload} event. When loading the HTML page is finished, the callback function is executed to dynamically fetch the root HTML of the target Web app and modify the DOM tree to render the actual page. Therefore, we can ensure that the retrieved HTML file of the target Web app is up-to-date.

To intercept resource requests, we use JavaScript reflection mechanism to register callback functions for all the cases of resource request. When the requested URL matches a URL's regular expression in the Resource Mapping, we replace the requested URL with the corresponding concrete URL to make the resource loaded from the AppCache.

Since the AppCache can work only in the next load after refresh, we explicitly call the \texttt{swap()} function of the AppCache when the AppCache's \emph{update} event is triggered in order to make the AppCache use the latest resources.

\subsection{Deployment}
Given the implementations based on the AppCache, developers can easily deploy ReWAP on their mobile Web architecture, requiring no extra cost to end users.

The whole ReWAP is deployed as a separate service on the Web server. Developers can configure the target mobile Web app to be integrated with ReWAP to launch a certain instance of ReWAP. When the ReWAP is launched, the Package Engine is automatically started as a background process on the server, while the Wrapper specific to the configured app is created in a certain folder. While the Package Engine is running, the two configuration files, Package Manifest and Resource Mapping, are also generated in the same folder as the Wrapper. Developers should make the folder accessible by the standard HTTP protocol where the Wrapper, Package Manifest and Resource Mapping are assigned dedicated URLs.

To make the Wrapper work for the target mobile Web app, the developer needs to only configure the server to redirect the entrance of the Web app to the URL of the Wrapper. Therefore, when users visit the ReWAP-enabled app, the Wrapper is first loaded to the browser. Nevertheless, the end-users are unaware of the existence of ReWAP when they request the target Web apps. 
\section{Evaluations}
\label{sec:evaluation}
We evaluate ReWAP from three main aspects\footnote{Please visit \url{http://taoxie.cs.illinois.edu/rewap/} to get more information of ReWAP implementation and evaluation results.}. First, we investigate the overall performance of ReWAP by measuring how much data traffic can be saved for mobile Web apps with ReWAP compared to the original apps that are with regular cache mechanisms enabled. Second, we evaluate the performance of the Package Engine, such as the resource normalization and the prediction of update time. Third, we evaluate the overhead of the Wrapper incurred to the mobile Web apps.

\subsection{Overall Performance}\label{sec:overall}
The performance of ReWAP is measured by how much data traffic can be saved compared to the original mobile Web apps. There are three main factors influencing the performance.

\noindent\textbf{Redundant transfers of the original mobile Web apps.} ReWAP aims to reduce redundant transfers for mobile Web apps. Therefore, if a mobile Web app has few redundant transfers, there may be not so much room for ReWAP to contribute to improve.

\noindent\textbf{User revisiting distribution.} Since ReWAP generates a unique resource package for all the users of a mobile Web app, we use a user revisiting distribution to calculate the benefit brought by different resource packages and choose the one with the largest benefit in our algorithm of generating resource packages (Algorithm~\ref{algo:select}). Certainly, different distributions could lead to different performances of ReWAP.

\noindent\textbf{Retrieving interval.} ReWAP helps reduce redundant transfers for retrieving resources of the same mobile Web apps. Generally, the longer the revisiting interval is, the lower the performance of ReWAP is because more resources may change after a longer interval, leading to the invalid of resource packages.

Given the preceding factors, we apply a simulation-based approach to evaluate the performance of ReWAP. We chose 50 mobile Web apps that suffer heavily from redundant transfers according to our previous measurement studies in WWW 2015 work~\cite{Ma:WWW2015}. Then we recorded all the resources of each mobile Web app's homepage every 30 minutes and lasted for 15 days. We assume that 30-minutes are sufficiently short to capture the changes of stable resources of Web apps. Based on this data set, we use the first five days' record to train our algorithms of normalization (Algorithm~\ref{algo:normalize}) and prediction (Algorithm~\ref{algo:predict}), and use the last ten days' records to generate resource packages. We evaluate the performance of ReWAP for revisiting intervals ranging from 0.5 hour to one day with 0.5 hour as an interval. Therefore, we have 48 revisiting intervals to investigate (0.5 hour, 1 hour, 1.5 hour, $\dots$, 24 hours). For user revisiting distributions, given a revisiting interval to investigate, we assume a 100\% revisiting of the same interval as we are to investigate. For comparison, we simulate an ideal cache with unlimited size to visit each Web app by each given revisiting interval. We calculate the differences of data traffic consumed by mobile Web apps with and without ReWAP. Note that the data traffic of ReWAP includes the cost of the update of resource packages.

\begin{figure}
\centering
\begin{center}
\includegraphics[width=1\textwidth,height=0.38\textwidth]{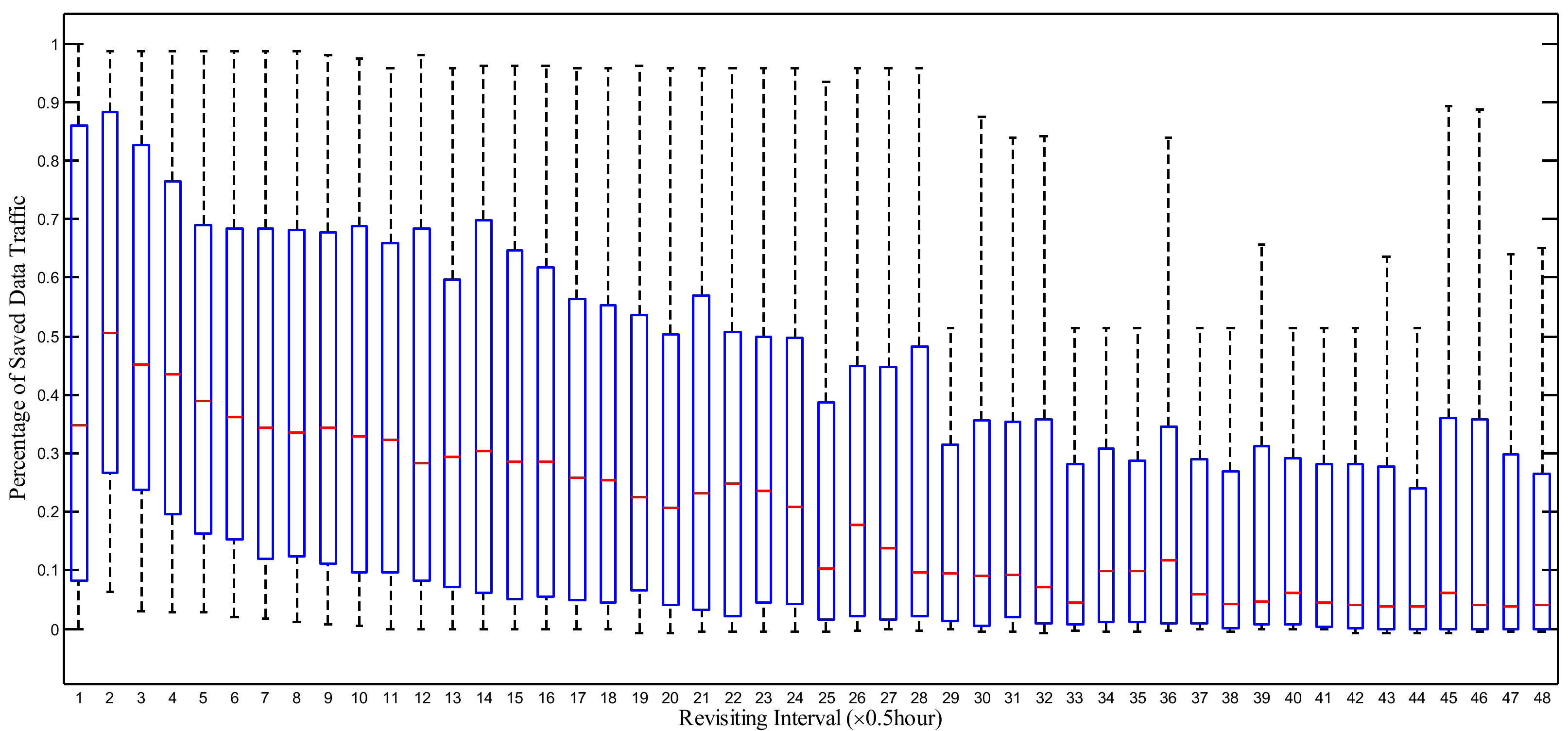}
\caption[7.5pt]{Distribution of saved data traffic comparing the case w/ and w/o ReWAP.}\label{fig:savedtraffic}
\end{center}
\end{figure}

Figure~\ref{fig:savedtraffic} shows the overall performance of ReWAP for different revisiting intervals. We draw a box plot to represent the distribution of saved data traffic for each revisiting interval. The median value of saved data traffic varies from 8\% to 51\%, indicating that mobile Web apps with ReWAP can reduce averagely 8\% to 51\% of the data traffic compared to the original Web apps with cache enabled. When the revisiting interval becomes larger, the saved data traffic decreases because the resource package has to be refreshed due to the update of packaged resources.

The variance of saved data traffic is significantly large for all revisiting intervals. In the best cases, the saved data traffic can reach almost 100\%, such as revisiting intervals shorter than 4 hours, implying that almost all the resources that should be downloaded from the network by the original Web app can be directly loaded from the local storage by ReWAP. In some worst cases especially for the longer revisiting interval, there is hardly improvement of achieved ReWAP because resources of some mobile Web apps are not stable enough to be packaged for longer revisiting intervals so the resources always change between two visits.

We should mention that more data traffic can be saved by ReWAP in the practice where the size of local cache is limited. The evaluation presented in this paper assumes an ideal cache where cached resources are always kept in the cache. However, in the real case, the size of cache is limited on mobile devices and cached resources are often removed out of the cache. Since ReWAP maintains an app-specific space for each Web app, the removed resources of original Web apps are also likely to be loaded from the local storage.

\subsection{Performance of the Package Engine}
The performance of ReWAP is determined by the resource packages generated by the Package Engine. More captured resources, more accurate predicted update time, and more stable resource packages can lead to saving more data traffic. We evaluate the performance of Package Engine by the intermediate data gathered during the experiment in Section~\ref{sec:overall}.

\begin{figure}[t]
\centering
\begin{center}
\includegraphics[width=1\textwidth,height=0.38\textwidth]{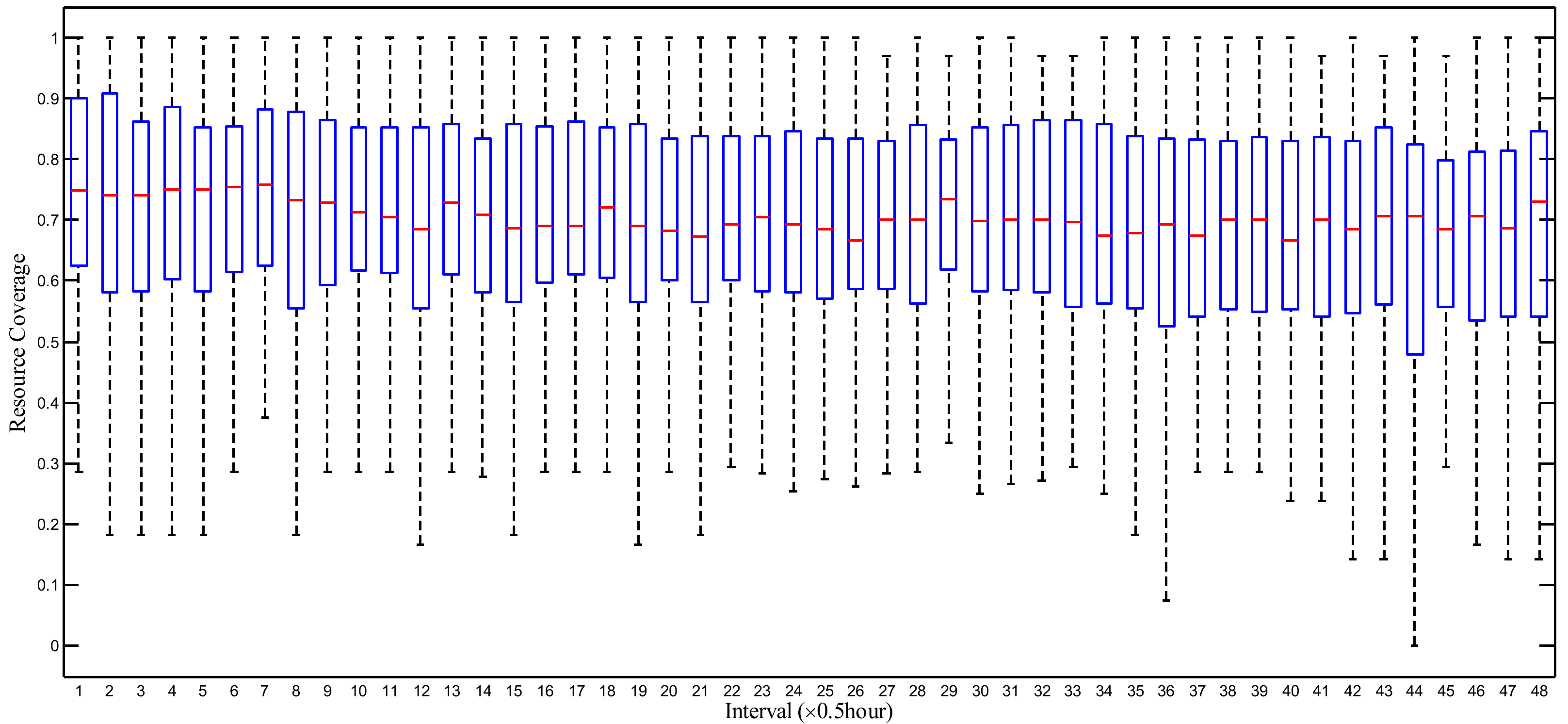}
\caption[7.5pt]{Distribution of the resource coverage among different intervals.}\label{fig:coverage}
\end{center}
\end{figure}

The Package Engine maintains a list of historic resources that are the candidates to be packaged. The list is changing every time when the Resource Retriever retrieves the new status of resources. It can be better if the resource list at a certain time $t$ covers more resources at the time later than $t$. Figure~\ref{fig:coverage} shows the distribution of the rate of resource coverage after different intervals ranging from 0.5 hour to one day. We can observe that the median coverage rate is around 70\% and it is very stable for different revisiting intervals. This result mainly accounts for the contribution of our normalization technique that cross-dress resources can be normalized to one resource. Therefore, more resources can be covered for longer durations.
\begin{figure*}[t]
\centering
\begin{center}
\includegraphics[width=1\textwidth,height=0.38\textwidth]{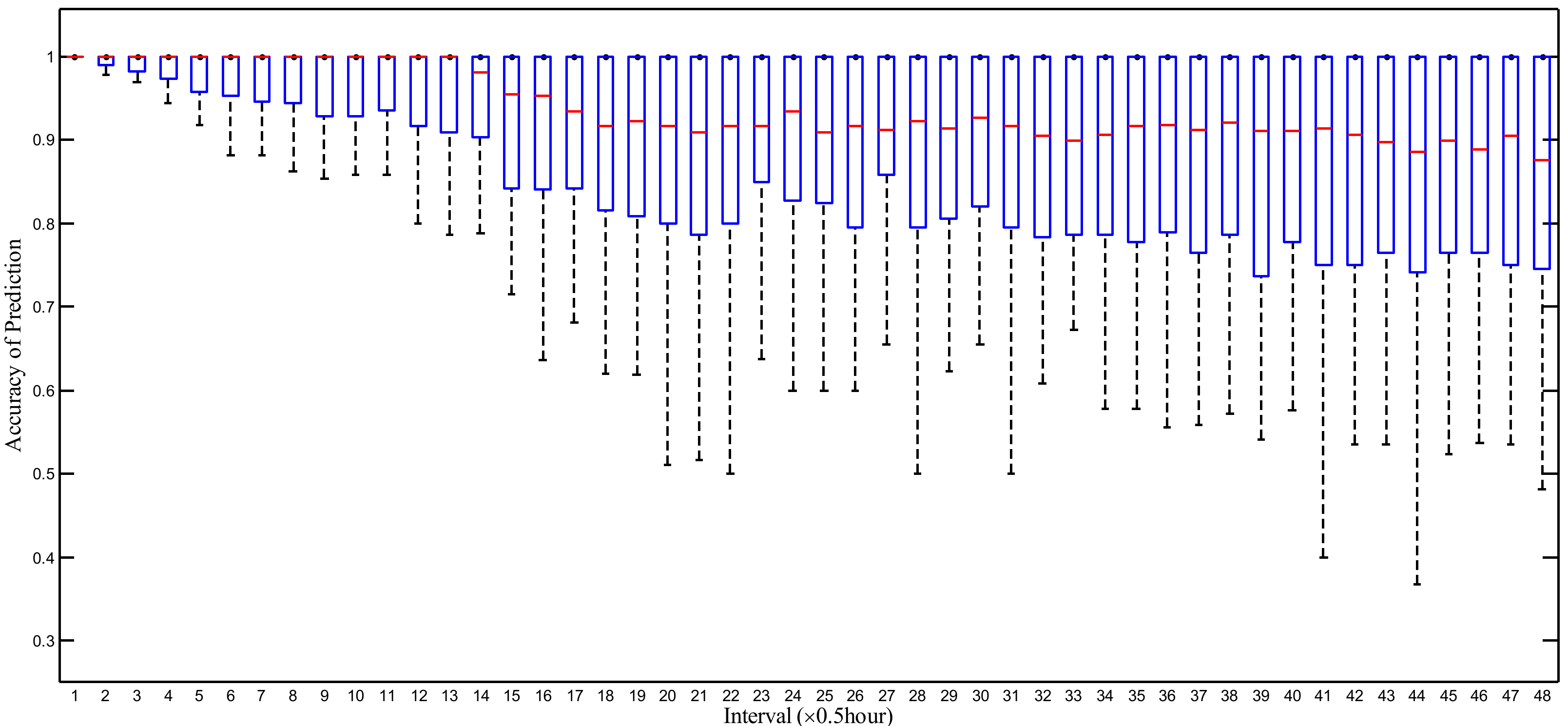}
\caption[7.5pt]{Distribution of the accuracy of predicted update time among different intervals.}\label{fig:accuracy}
\end{center}
\end{figure*}
The Package Engine uses an estimated update time to judge whether a resource is stable. Since the update time is an important factor in calculating the benefit of resource packages in Algorithm~\ref{algo:predict}, the predictions need to be precise enough. Figure~\ref{fig:accuracy} shows the distribution of the precision of predicted update time. We can observe that the precision decreases as the interval increases. For all the intervals, the median predicting precision is above 85\%. For intervals less than 5 hours, the median precision is 100\%. Overall, such accuracy can be satisfying to most apps and demonstrates the effectiveness of ReWAP.
\begin{figure}
\centering
\begin{center}
\includegraphics[width=0.45\textwidth]{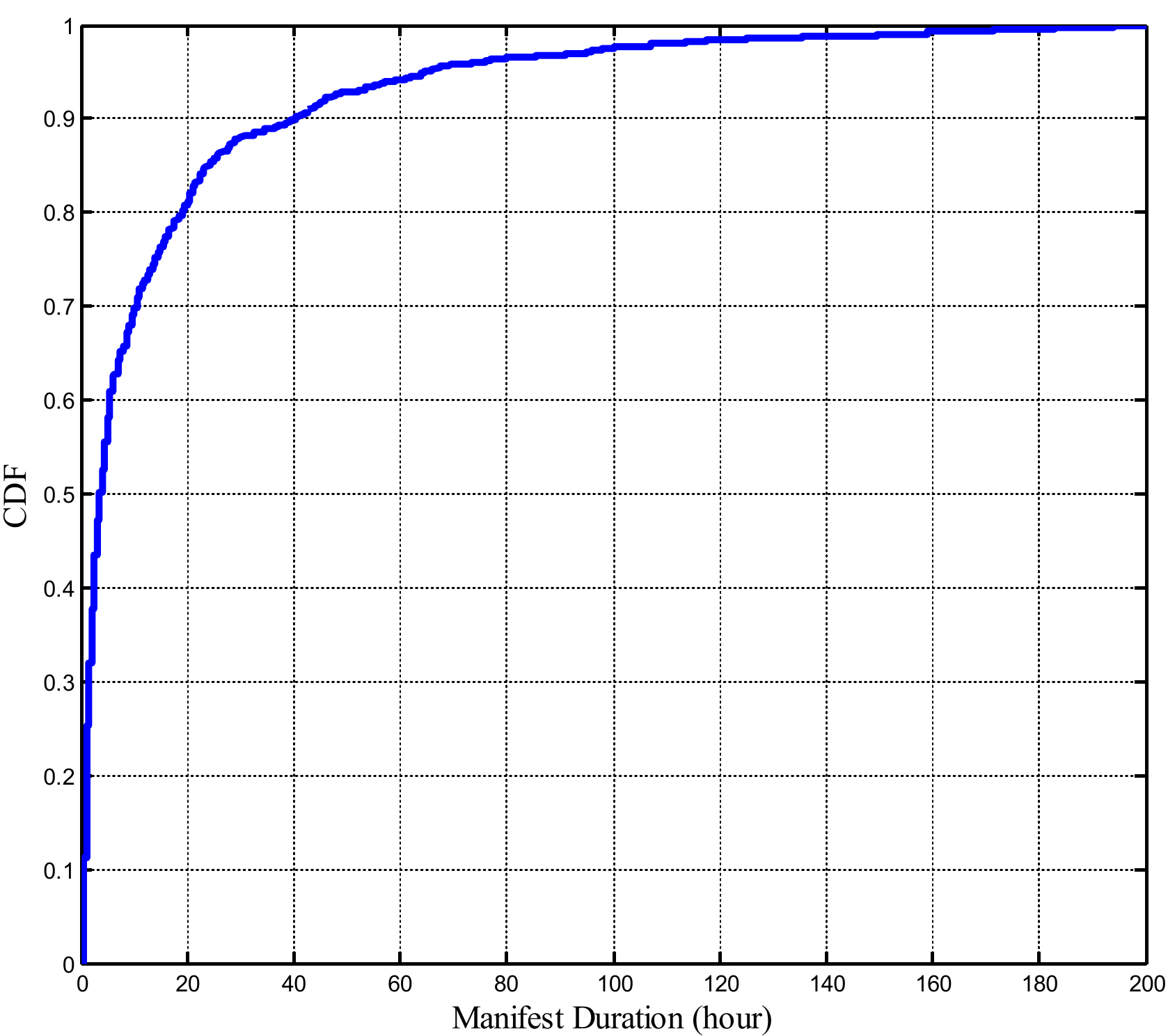}
\caption[7.5pt]{Distribution of the durations of resource package.}\label{fig:manifestduration}
\end{center}
\end{figure}
The more stable the resource packages are, the less refresh the Wrapper performs. We calculate the time duration between every two updates of resource packages. Figure~\ref{fig:manifestduration} shows the distribution of the duration. We can observe that the median duration of a resource package is 5 hours, indicating that resource packages should be updated every 5 hours in the medium cases. Therefore, the performance of ReWAP is better for revisiting intervals less than 5 hours in the situation of the experiment in Section~\ref{sec:overall}.

\subsection{Overhead of the Wrapper}
When a ReWAP-enabled mobile Web app runs on the browsers, the Wrapper can introduce overhead compared to the original app. The overhead lies in two main aspects. One is the manifest files to specify resource packages. The other is the computation logics of resource mapping and the Application Cache itself.

\begin{figure}
\centering
\begin{center}
\includegraphics[width=0.45\textwidth]{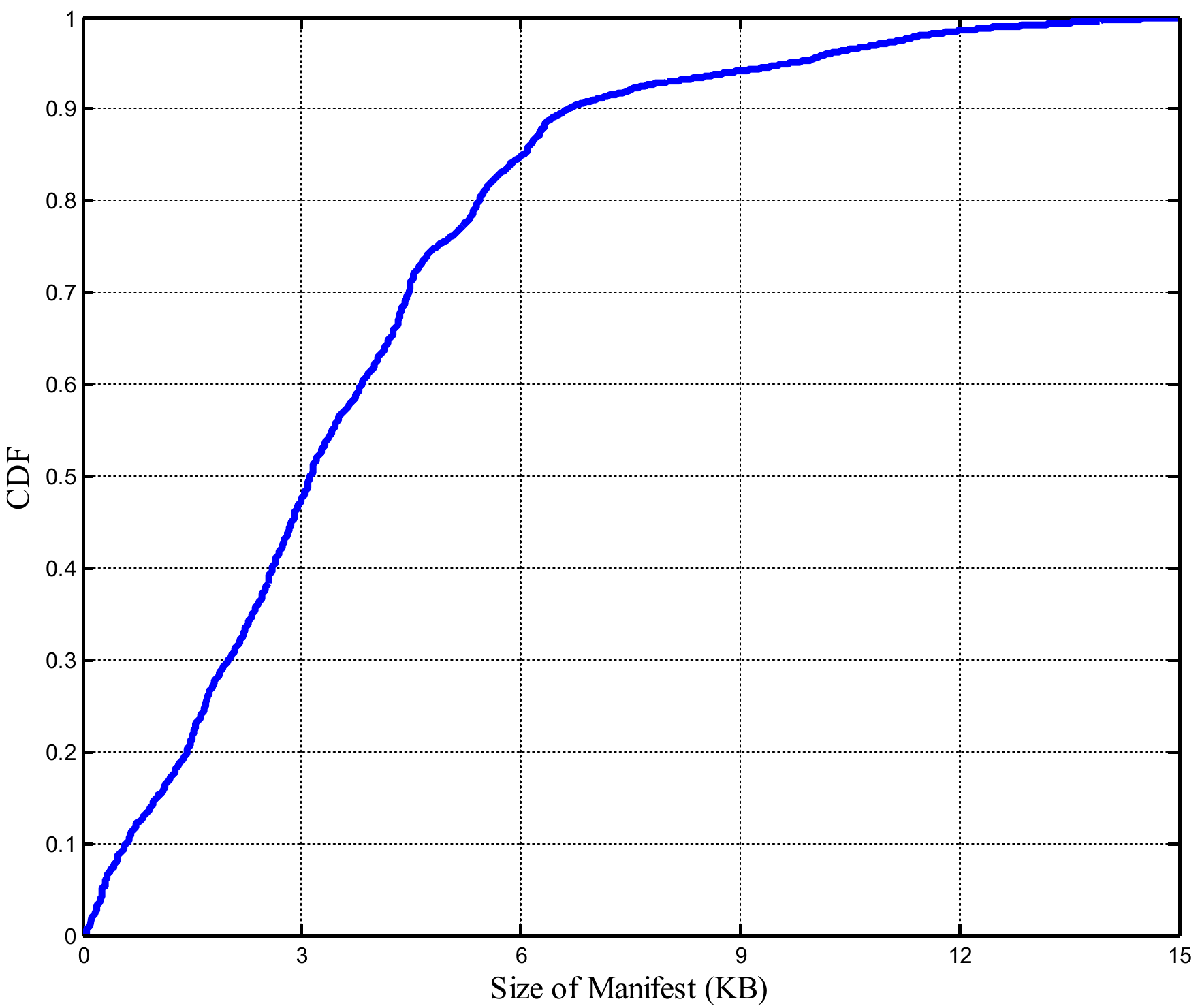}
\caption[7.5pt]{Distribution of the size of Package Manifest.}\label{fig:manifestsize}
\end{center}
\end{figure}

Figure~\ref{fig:manifestsize} shows the distribution of manifest files' size, which is gathered from the manifests generated in the experiment of Section~\ref{sec:overall}. We can see that the median size is only 5 KB and the largest is not more than 20 KB. Therefore, the overhead of manifest file is small enough.

To evaluate the overhead of computation logics and their influences on the page load time, we generate some test pages whose number of resources ranges from 20 to 100, and each resource is 100 KB. We assume that all the resources of each page are put into the corresponding package and 10\% of the packaged resources are normalized resources that are identified by URL regular expressions. With the assumption, we generate the manifest files. Then we visit each page twice in the browser on a smartphone of Samsung Galaxy S4 with Android 5.0. We record the CPU usage and memory usage as well as the page load time during loading each page with and without ReWAP. We find that the average CPU usage is increased by 15\% for pages with ReWAP while the memory usage is of no significant differences.

\begin{figure}
\centering
\begin{center}
\includegraphics[width=0.45\textwidth]{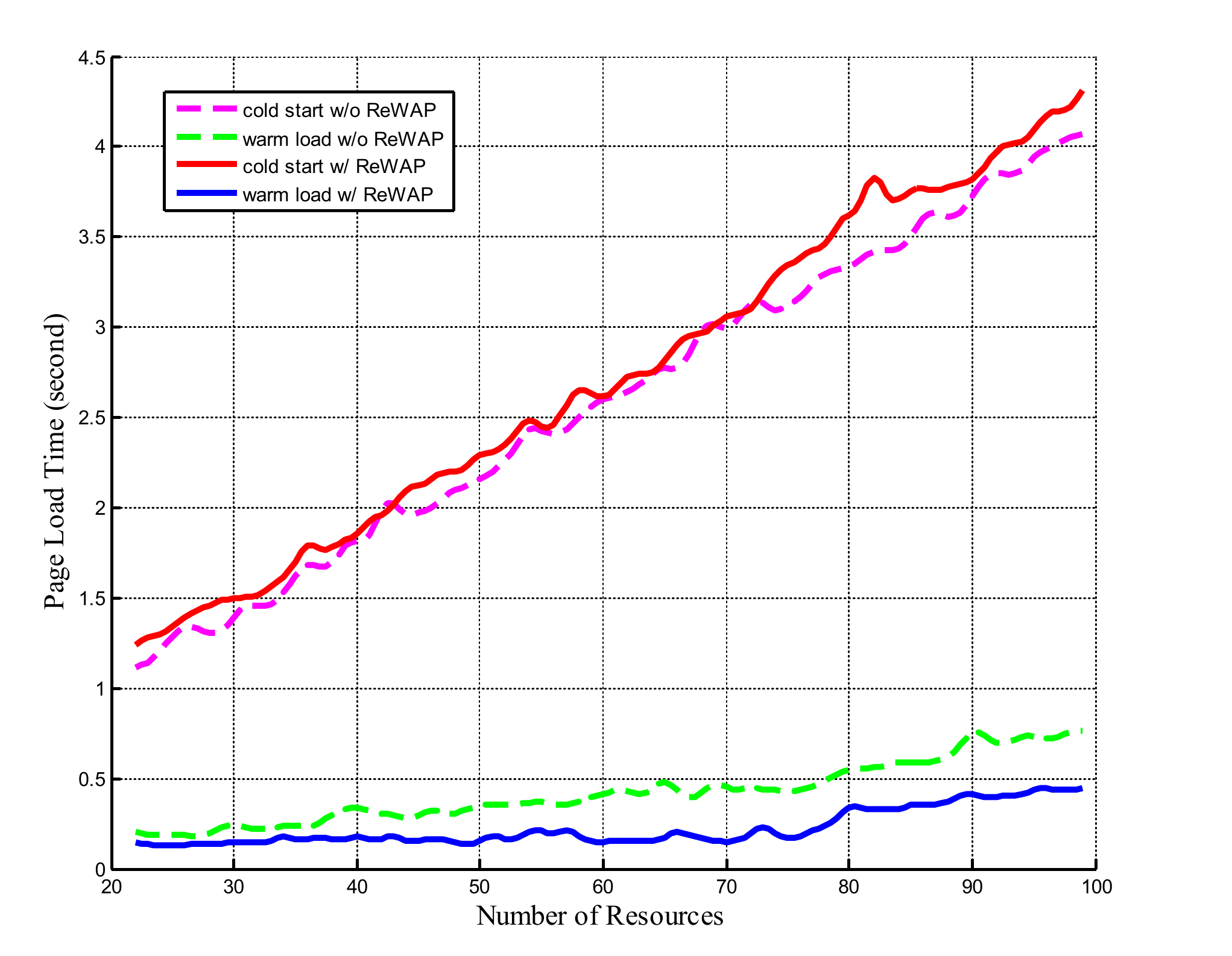}
\caption[7.5pt]{Influences of ReWAP on Page load time for cold start and warm load.}\label{fig:loadtime}
\end{center}
\end{figure}

Figure~\ref{fig:loadtime} shows the page load time for different pages in four cases: cold start without ReWAP, cold start with ReWAP, warm load without ReWAP, and warm load with ReWAP. We can observe that as the number of resources increases, the page load time increases in all cases, and the gap between cold start and warm load becomes bigger no matter whether the app is equipped with or without ReWAP. For the cold start, the page load time of pages with ReWAP is a little longer than those without ReWAP. However, for the warm load, the page time of pages with ReWAP is much shorter than those without ReWAP. This observation implies that ReWAP can also reduce the page load time when mobile Web apps are revisited. In the best cases, the page load time is reduced by more than 50\%.

In summary, we can conclude that the overhead of the Wrapper is very minor. ReWAP can even reduce the page load time when mobile Web apps are revisited. Therefore, the results demonstrate that ReWAP is practical to be adopted and applied.

\section{Related Work}
\label{sec:related}
It is well known that the user experiences of mobile Web apps are far from satisfaction in terms of the page load time, data traffic drain, and energy consumption. Redundant transfers of resources are the most dominant issue leading to such inefficiency. Our work focuses on reducing redundant transfers to improve the user experiences of mobile Web apps. We then discuss related work.

\textbf{Measurement studies on resource loadings of mobile Web apps.} Wang et al.~\cite{Wang:HotMobile2011} advocated that resource loading contributes most to the browser delay. Wang et al.~\cite{Wang:NSDI2013} designed a lightweight in-browser profiler, called WProf, and studied the dependencies of activities when browsers load a webpage. Nejati et al.~\cite{Nejati:WWW2016} extended WProf to WProf-M and studied the differences of page loading process between mobile and non-mobile browsers. Li et al.~\cite{Li:NSDI2010} designed WebProphet to capture dependencies among Web resources and to automate the prediction of user-perceived Web performance. The poor performance of mobile Web cache is a key issue leading to redundant transfers. Qian et al.~\cite{Qian:MobiSys2012} measured the performance of mobile Web cache in terms of the cache implementation and revealed that about 20\% of the total Web traffic examined is redundant due to imperfect cache implementations. In their later work~\cite{Qian:MobiSys2014}, they studied the caching efficiency for the most popular 500 websites and found that caching is poorly utilized for many mobile sites. Wang et al.~\cite{Wang:WWW2012} found that cache has very limited effectiveness: 60\% of the requested resources are either expired or not in the cache. Our previous work~\cite{Ma:WWW2015} adopted a proactive approach to measure the performance of mobile Web cache and found that more than 50\% of resource requests are redundant on average for 55 popular mobile websites. In particular, we found some underlying factors leading to redundant transfers of mobile Web apps, i.e., Same Content, Heuristic Expiration, and Conservative Expiration Time. These measurement studies motivate us to reduce redundant transfers for mobile Web apps.

\textbf{Techniques to reduce redundant transfers.} Wang et al.~\cite{Wang:IMC2014} investigated how Web browsing can benefit from micro-cache that separately caches layout, code, and data at a fine granularity. They studied how and when these resources are updated, and found that the layout and code that block subsequent object loads are highly cacheable. Our resource packaging can be viewed as to realize similar features proposed by micro-cache in the resource granularity. Most of the stable layout and code resources are put into the package to always be loaded from the local environment. Zhang et al.~\cite{Zhang:UbiComp2013} implemented a system-wide service called CacheKeeper, to effectively reduce overhead caused by poor Web caching of mobile apps. CacheKeeper can also work for browsers but it relies on the support of operating systems. Our implementation of ReWAP utilizes the standard HTML5, which has been supported by all the commodity mobile Web browsers. So we can achieve fast and easy deployment with no cost to end users.

\textbf{General techniques to improve the performance of mobile Web.} Some previous work focuses on improving the compute-intensive operations for mobile Web browsers, such as style formatting~\cite{Wang:WWW2014}, layout calculation~\cite{Zhang:WWW2010,Mazinanian:FSE2014}, and JavaScript execution~\cite{Huang:MobiSys2010,Gong:FSE2015}. However, Wang et al.~\cite{Wang:HotMobile2011} argued that the key to improve the performance of mobile Web is to speed up resource loading. Various solutions have been proposed to optimize resource loading. These solutions include new network protocols such as SPDY~\cite{Wang:NSDI2014} and HTTP2~\cite{HTTP2}, browser optimization such as prefetching~\cite{Lymberopoulos:ASPLOS2012} and speculative loading~\cite{Wang:WWW2012}, and proxy-based systems such as Flywheel~\cite{Agababov:NSDI2015} and KLOTSKI~\cite{Butkiewicz:NSDIS15}. These previous solutions are orthogonal to reducing redundant transfers.

Dynamics and user revisits of Web pages. Douglis et al.~\cite{Douglis:USITS1997} performed a live study on the influences of resource changes and user revisits on the Web caching in early 1997. Fetterly et al.~\cite{Fetterly:WWW2003} measured the degree of Web page changes and investigated the factors correlated with change intensity. Adar et al.~\cite{Adar:CHI2008} studied the Web revisiting behaviors from a live data set. They identified four revisiting patterns for different kinds of Web pages. In their later work~\cite{Adar:CHI2009}, they studied the relationship between the dynamics of Web pages and user revisiting patterns. Although all these previous efforts focus on the Web for desktop computers, their findings can be partly leveraged by the mobile Web. Our work depends on the dynamics and user revisits of Web pages to maintain the resource package.
\section{Conclusion}
\label{sec:conclusion}
In this paper, we have presented the ReWAP approach, by radically making the resource management of mobile Web apps perform in a similar fashion to the native apps. The key rationale of ReWAP is to provide more efficient and ``application-aware'' control of resource management rather than relying on only the current mechanisms such as Web cache, to avoid the caused unnecessary redundant resource transfer. To realize the efficient resource packaging, our ReWAP includes a normalization technique to identify the same resources but with different URLs, a learning-based technique to accurately predict the updates of resources, and an algorithm to minimize the refresh frequency of resource packages as well as reduce the overhead. We have evaluated ReWAP based on long-term (15-day) traces of existing Web apps that suffer from redundant resource transfer, and the results demonstrate our approach's effectiveness and efficiency.

Given that ReWAP can be easily deployed into existing Web applications with very few manual efforts, our ongoing work is to encapsulate ReWAP as an independent module into currently popular Web servers such as Apache, Nginx, and Node.js. When ReWAP is deployed and the actual access logs are obtained, we can derive the distribution of user's visit frequency for a given Web app, and thus optimize the prediction algorithm by tuning the parameters and designing online learning kernels.

While our work in this paper addresses the key challenges caused by the dynamics of Web apps, some other issues are worth further exploring. For example, although the data traffic can be significantly reduced, it is observed that the page load time of Web apps does not always improve due to the checking update with server-side components.  Prefetching the updates of resource packages at a desirable time or the ¡°push-oriented¡± update notification can be potential solutions. Other performance issues, e.g., the energy drain, need to be addressed as well. 

\bibliography{webos}
\bibliographystyle{IEEEtran}

\end{document}